\documentclass[aip,amsmath,amssymb,reprint, twocolumn]{revtex4-1}
\usepackage[utf8]{inputenc}
\usepackage[T1]{fontenc}
\usepackage{mathptmx}
\usepackage{rotating}
\usepackage{longtable}
\usepackage{listings}
\usepackage{xcolor}
\usepackage{siunitx}
\usepackage{lipsum}
\usepackage{tikz}
\usetikzlibrary{arrows.meta}
\usepackage{wrapfig}
\definecolor{codegreen}{rgb}{0,0.6,0}
\definecolor{codegray}{rgb}{0.5,0.5,0.5}
\definecolor{codepurple}{rgb}{0.58,0,0.82}
\definecolor{backcolour}{rgb}{0.95,0.95,0.92}
\lstdefinestyle{mystyle}{
    backgroundcolor=\color{backcolour},   
    commentstyle=\color{codegreen},
    keywordstyle=\color{magenta},
    numberstyle=\tiny\color{codegray},
    stringstyle=\color{codepurple},
    basicstyle=\ttfamily\footnotesize,
    breakatwhitespace=false,         
    breaklines=true,                 
    captionpos=b,                    
    keepspaces=true,                 
    numbersep=5pt,                  
    showspaces=false,                
    showstringspaces=false,
    showtabs=false,                  
    tabsize=2
}
\PassOptionsToPackage{hyphens}{url}
\usepackage{xurl}
\usepackage{hyperref}
\hypersetup{
  colorlinks=true,
  citecolor=blue,
  filecolor=blue,
  linkcolor=blue,
  urlcolor=blue,
  breaklinks=true
}

\usepackage{times,amsmath}
\usepackage{epsfig}
\usepackage{color}
\usepackage{longtable}
\usepackage{ulem}
\usepackage{graphicx}
\usepackage{dcolumn}
\usepackage{bm}
\usepackage{bookmark}
\usepackage{tabularx}
\usepackage{multirow}
\usepackage{booktabs}
\usepackage{tocloft}
\usepackage{etoolbox}
\apptocmd{\thebibliography}{\sloppy}{}{}
\usepackage[switch, pagewise]{lineno}
\linenumbers\relax 
\nolinenumbers

\begin{document}

\title{Ab-initio Crystal Structure Determination from Powder X-Ray Diffraction}

\author{Kaixiang Su}
\affiliation{Department of Computer Science, University of North Carolina at Charlotte, Charlotte, NC 28223, USA}

\author{Osman Goni Ridwan}
\affiliation{Department of Mechanical Engineering and Engineering Science, University of North Carolina at Charlotte, Charlotte, NC 28223, USA}

\author{Hongfei Xue$^*$}
\email{hongfei.xue@charlotte.edu}
\affiliation{Department of Computer Science, University of North Carolina at Charlotte, Charlotte, NC 28223, USA}

\author{Qiang Zhu$^*$}
\email{qzhu8@charlotte.edu}
\affiliation{Department of Mechanical Engineering and Engineering Science, University of North Carolina at Charlotte, Charlotte, NC 28223, USA}
\affiliation{North Carolina Battery Complexity, Autonomous Vehicle and Electrification (BATT CAVE) Research Center, Charlotte, NC 28223, USA}

\date{\today}
\begin{abstract}
\textbf{Abstract.}
Determining crystal structures from powder X-ray diffraction (PXRD) has been a significant challenge in materials science, particularly when experimental data contain noise or the target structure has a high complexity. While recent AI generative models show promise for rapid structure generation, they predominantly employ data-driven approaches to learn direct mappings between PXRD patterns and crystal structures, often failing on complex or out-of-distribution cases. To overcome these limitations, we present a hybrid ab-initio approach that decomposes structure determination into a two-stage optimization problem: (1) discrete selection of space group symmetry, unit cell parameters, and Wyckoff site combinations; and (2) continuous optimization of atomic coordinates within the selected Wyckoff positions. By integrating AI-based techniques for peak profile analysis, density estimation and energy minimization with physics-informed constraints, our method systematically overcomes limitations of purely data-driven PXRD solvers. We demonstrate that this physics-AI hybrid optimization framework enables robust structure determination across a diverse benchmark of over 1,000 challenging crystal structures where state-of-the-art generative PXRD solvers fail. This work provides a scalable, principled pathway for combining domain-specific crystallographic knowledge with modern machine learning to achieve reliable and generalizable structure solution from experimental data.

\end{abstract}

\maketitle
\makeatletter

\vspace{3mm}\noindent
\textbf{\large{1. Introduction}}\\

In materials research, experimental data obtained from techniques such as powder X-ray diffraction (PXRD) and electron microscopy are often incomplete or ambiguous, making direct structure solution difficult. Deciphering structure from such data is a complex inverse problem that requires advanced computational methods to identify plausible solutions consistent with experimental observations. Traditional approaches typically generate a pool of random candidate structures based on experimental data, followed by refinement through minimization techniques, which becomes highly challenging for complex systems. Recent advances in artificial intelligence (AI) generative models and machine learning (ML) have opened new possibilities for addressing this challenge~\cite{de2025generative}. These models, including GANs~\cite{kim2020generative, zhu2024wycryst}, VAEs~\cite{noh2019inverse, ren2022invertible,  ridwan2025ai}, diffusion models~\cite{CDVAE, DiffCSP, diffcsp++, levy2024symmcd, zeni2025generative}, flow models~\cite{miller2024flowmm}, transformers~\cite{de2025generative-w, cao2024space, antunes2024crystal}, and large language models (LLMs)~\cite{gruver2024fine}, provide a fundamentally new approach to exploring structure space by learning data-driven representations from chemical formulas, enabling efficient generation of low-energy crystal structures at a much faster rate than traditional global optimization methods. Building on these models, many recent studies incorporate PXRD information as an additional input for conditional generation to solve experimentally observed crystal structures
\cite{oviedo2019fast, aguiar2019decoding,
venderley2022harnessing, yanxon2023artifact, chen2024crystal, maffettone2021crystallography, choudhary2025diffractgpt, guo2025ab, zhao2025xdxdendtoendcrystalstructure,johansen2026decifercrystalstructureprediction, cao2025simxrdm, li2025powder, lu2026unified, larsen2024phai, guan2026deep}. 
These approaches aim to directly learn additional mapping from PXRD patterns to crystal structures, holding promise for more rapid and accurate structure determination.

Although these generative models can quickly produce crystal structures, it remains unclear whether they can adequately learn and enforce the physical constraints encoded in PXRD data when inferring new structures. From the X-ray diffraction perspective, PXRD patterns contain two primary sources of structural information: unit-cell parameters inferred from peak positions and atomic arrangements reflected in peak intensities.
As shown in Table \ref{tab:stat1}, adapted from a recent survey~\cite{li2025powder}, incorporating PXRD information yields only marginal improvements in the top-20 match rates of diffusion and flow models and still does not recover the accuracy achieved when unit-cell parameters are known. Even when cell parameters are provided, the best reported accuracy (87.68\%--87.97\%) indicates that it remains hard to decipher the atomic coordinates correctly for over 12\% of material systems within the commonly used MP-20 data set~\cite{MP-2013, CDVAE}. Taken together, these observations suggest that current generative models are not effectively learning the underlying physics of PXRD patterns and may instead be capturing superficial correlations in the training data.

\begin{table}[ht]
    \caption{Top-20 match rates (\%) for crystal structure determination from PXRD. Results are adapted from Li et al.~\cite{li2025powder}.}
    \centering
    \begin{tabular}{lc}
    \hline
    \textbf{Model}     &  \textbf{Top-20 Match Rate (\%)}\\
    \hline
    Base Diffusion                 &  77.69\\
    Diffusion + PXRD encoder       &  83.87\\
    Base Diffusion + Cell Parameters    &  87.68\\
    Base Flow                      &  80.23\\
    Flow + PXRD encoder            &  85.37\\
    Base Flow + Cell Parameters         &  87.97\\
    \hline
    \end{tabular}
    \label{tab:stat1}
\end{table}

\begin{figure*}[htbp]
    \centering
    \vspace{-2mm}
    \includegraphics[width=0.93\textwidth]{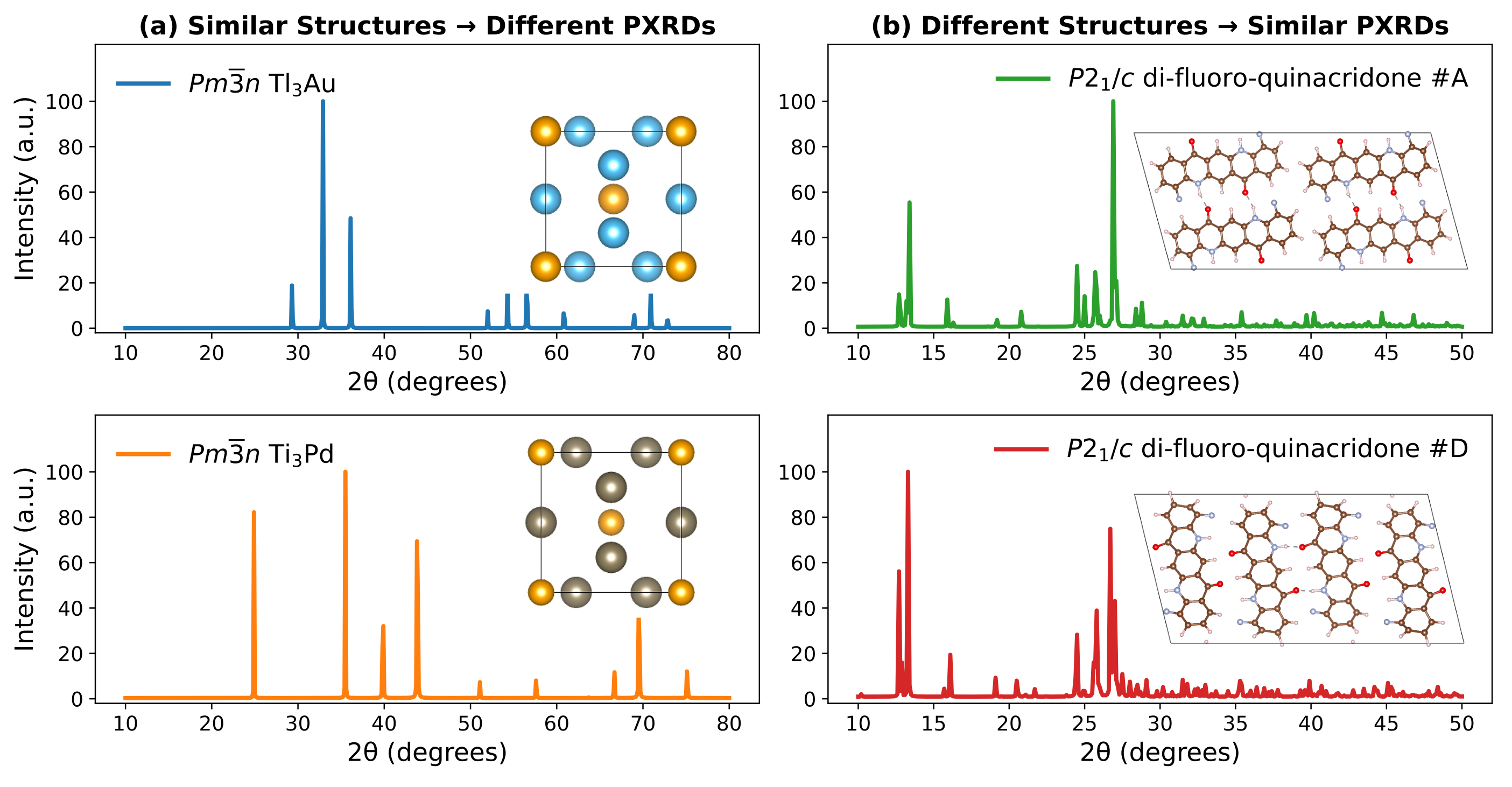}
    \vspace{-3mm}
    \caption{\small{The mapping challenge between crystal structure and measured PXRD. (a) Two isostructural $Pm\overline{3}n$ Tl$_3$Au and Ti$_3$Pd exhibit different PXRD patterns due to complex interplay between atomic scatterings. (b) Two structures (mostly with low symmetry) may exhibit very similar PXRD patterns despite the structural dissimilarity \cite{schlesinger2022ambiguous}.}}
    \label{fig:pxrd}
    \vspace{-3mm}
\end{figure*}   

Furthermore, our benchmarks on two recently developed models applied to the MP-20 dataset reveal a systematic overlap in failure cases. In Table \ref{tab:stat2}, of the 1318 failures produced by PXRDGen \cite{li2025powder} and the 1817 failures produced by Uni-3DAR \cite{lu2026unified}, 1156 are shared between the two models. This striking overlap strongly suggests that there may exist a fundamental obstacle to prevent the mapping between PXRD and crystal structure via statistical approach. 

\begin{table}[ht]
    \caption{Failure comparison between two generative models on the MP20 dataset according to the top-20 predictions.}
    \vspace{2mm}
    \centering
    \begin{tabular}{lcc}
    \hline
    \textbf{Model}     &  \textbf{Total Failures} & \textbf{Shared Failures}\\
    \hline
    PXRDGen \cite{li2025powder}    &  1318 & 1156\\
    Uni-3DAR \cite{lu2026unified}  & 1817  & 1156\\
    \hline
    \end{tabular}
    \label{tab:stat2}
\end{table}

From a physics perspective, the root cause of these failures likely lies in the fact that a direct PXRD spectrum is neither a smooth nor a unique representation of crystal structure (Fig.~\ref{fig:pxrd}). Even two isostructural compounds can exhibit different PXRD patterns (Fig.~\ref{fig:pxrd}a). Conversely, two structurally different crystals may produce very similar PXRD patterns (Fig.~\ref{fig:pxrd}b), leading to nearly indistinguishable metrics in subsequent post-refinement analysis~\cite{schlesinger2022ambiguous}. This intrinsic complexity makes establishing a reliable one-to-one mapping between structure and PXRD inherently difficult. Although modern AI methods can drive models to learn smooth latent spaces from existing data~\cite{guo2025ab,li2025powder}, this relationship often generalizes poorly to unseen data because of the underlying non-uniqueness.

\begin{figure*}[htbp]
    \centering
    \vspace{-2mm}
\includegraphics[width=0.99\textwidth]{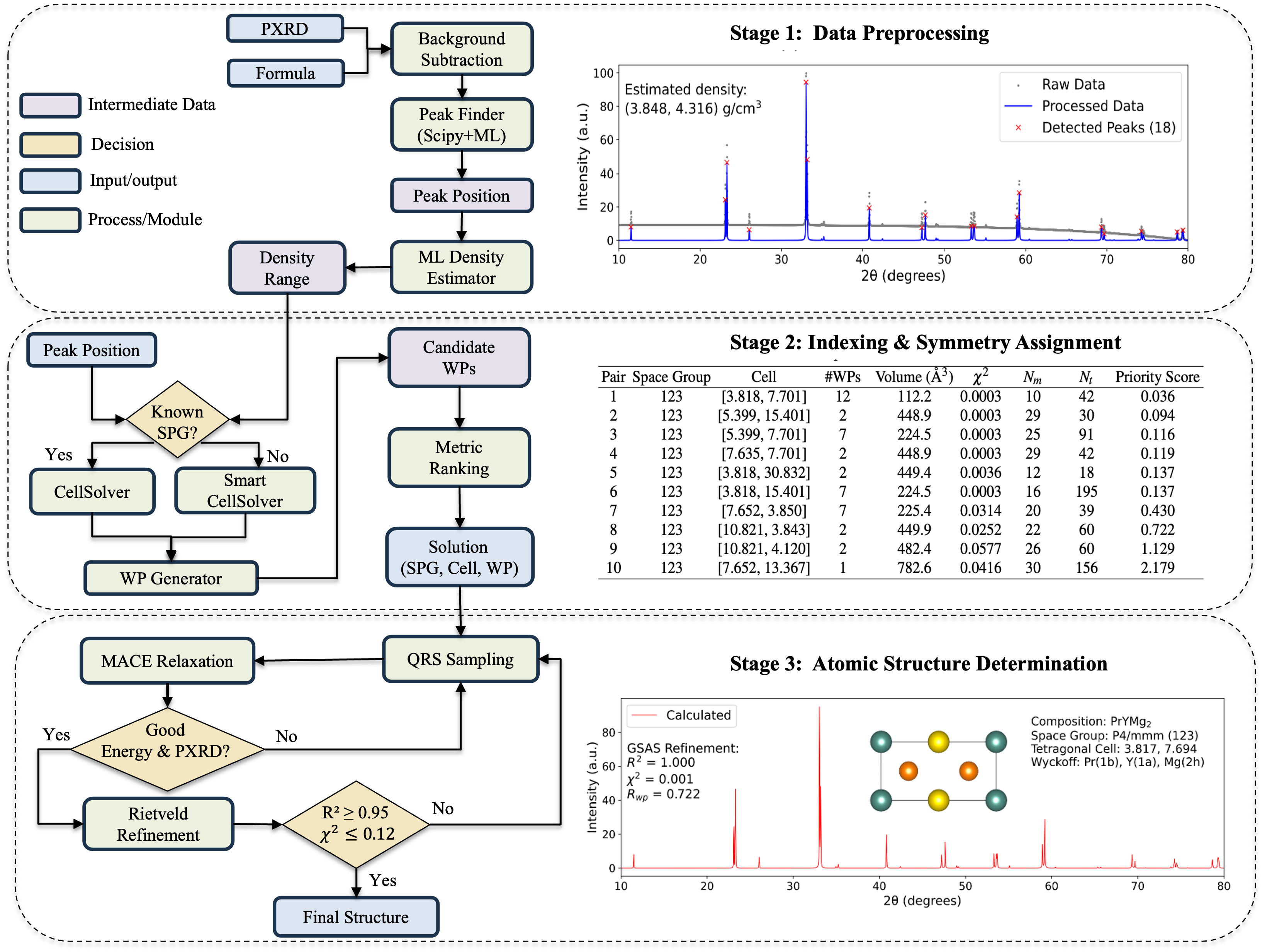}
    \vspace{-3mm}
    \caption{The outline of Ab-PXRD-Solver pipeline based on three stages model.}
    \label{fig:workflow}
    \vspace{-1mm}
\end{figure*}

Therefore, we emphasize that solving crystal structures from PXRD is a daunting task that requires extracting physically meaningful information from diffraction patterns while managing additional complexity from measurement uncertainty. Although a fast end-to-end solver is desirable, materials researchers can typically tolerate solving complex structures within a few days rather than a few seconds. Hence, accuracy is more critical for practical applications. We therefore combine relevant physical knowledge with emerging AI/ML techniques throughout the solution process. In this work, we present an \textit{ab initio} PXRD solver approach that decomposes structure determination into a sequence of optimization subproblems and selects appropriate solvers for each subproblem based on the task nature, using either physics-based or AI-based techniques. By leveraging the strengths of both approaches, we can effectively integrate AI/ML modules with physically grounded sub-solvers. Using this combined solver, we demonstrate substantially improved robustness and reliability. In the following sections, we describe the proposed pipeline in detail, illustrate its performance through representative examples, present a large-scale benchmark, and discuss future directions.

\vspace{3mm}\noindent
\textbf{\large{2. The Ab-PXRD-Solver Workflow}}\\

As summarized in Fig.~\ref{fig:workflow}, our Ab-PXRD-Solver pipeline decomposes structure determination into three sequential stages: (1) data preprocessing for PXRD peak management and density estimation; (2) cell indexing to identify unit cell parameters and symmetry; and (3) atomic structure determination via relaxation and post-refinement.

\vspace{2mm}\noindent
\textbf{Stage 1: Data Pre-processing.}
In a standard structure-determination task, the PXRD pattern and chemical formula are typically known. To obtain a feasible structural solution, two pieces of critical information must be extracted from these inputs. First, unique PXRD peak information (including both position and intensity) must be identified. In real experimental settings, raw PXRD data often contain substantial noise and background signals, which can lead to false peak detection and, consequently, incorrect structural solutions. Second, explicit information about the material density is usually unavailable, even though it is critical for constraining the search space of possible structures.

\vspace{2mm}\noindent
\textit{Stage 1.1: Background Subtraction and Peak Extraction}. To extract reliable PXRD peak information from raw data, we apply a sequence of signal-processing and machine-learning methods. First, we perform adaptive background subtraction using asymmetric least-squares polynomial fitting, followed by smoothing to suppress noise while preserving peak shapes. Peak candidates are initially detected using SciPy's peak finder with a minimum normalized height threshold (e.g., 0.02). Although this approach is effective for strong peaks, it may include artificial shoulders caused by peak broadening. To address this, we further filter candidates using a pretrained convolutional neural network (CNN) that predicts peak probability. A candidate is retained unless its predicted probability falls below 0.8 and its intensity is below a height threshold, thereby minimizing false negatives. Optional expert intervention is recommended when instrumentation artifacts are suspected. The upper right plot of Fig.~\ref{fig:workflow} shows the results of background subtraction and peak extraction.

\vspace{2mm}\noindent
\textit{Stage 1.2 Density Estimation}. We leverage a pretrained message-passing neural network ensemble~\cite{goodall2020predicting} to predict the density bounds ($\rho_{\min},~\rho_{\max}$) from the given chemical composition. This density estimation provides a crucial constraint on the physically admissible unit-cell volume for subsequent stages of cell indexing and Wyckoff-site assignment, thereby improving both the efficiency and accuracy of structure determination.

\vspace{2mm}\noindent
\textbf{Stage 2: Unit Cell Indexing and Symmetry Assignment.} Once the PXRD peaks and material density are available, we index the peaks according to Bragg's law to determine the space-group symmetry and unit-cell parameters, and then generate possible Wyckoff sites that are compatible with the density constraints. At this stage, the space-group number can be provided either explicitly or inferred from the data.

\vspace{2mm}\noindent
\textit{Stage 2.1 Solving the Cell from the Known Space Group.} When the space-group is known, \textit{CellSolver} determines the six cell parameters ($a, b, c, \alpha, \beta, \gamma$) by enumerating symmetry-allowed $(hkl)$ triples up to a configurable maximum and constructing a series of linear equations based on Bragg's law. Depending on the crystal symmetry, the required number of independent $(hkl)$ planes ($N_{hkl}$) ranges from 1 for cubic systems to 6 for triclinic systems. Accordingly, we select $N_{hkl}$ nonzero $(hkl)$ candidates from the first few peaks under the given SPG and derive the cell parameters by solving the resulting linear equations (see Supplementary Materials). To avoid linearly dependent $(hkl)$ sets, we additionally include 2--5 extra peaks in this mapping step. Once the cell parameters are obtained, they are used to index the remaining peaks. In general, a good solution yields calculated peaks that match all observed peaks within a numerical threshold, as quantified by the residual error \(\chi^2_\text{cell} = \frac{1}{N \sigma^2}\sum_{i=1}^{N}\left(\theta_i^{\mathrm{obs}} - \theta_i^{\mathrm{calc}}\right)^2\), where $i$ runs over all matched peaks, while producing only a few predicted $(hkl)$ peaks within a short 2$\theta$ range that are not observed experimentally. We denote the number of such unmatched peaks by $N_m$. These missing peaks may arise from noise, peak overlap, or diffraction cancellation caused by atomic arrangements. A supercell can also yield a low $\chi^2_\text{cell}$ but a large $N_m$ because of the many additional reflections. We therefore evaluate each trial solution jointly using $\chi^2_\text{cell}$ and $N_m$.
Many space groups also obey distinct extinction rules imposed by crystallographic symmetry~\cite{Aroyo2016ITA}. We leverage the extinction rules for all 230 space groups through PyXtal's symmetry module~\cite{pyxtal,zhu2022symmetry} to filter $(hkl)$ candidates for each SPG, thereby improving both the efficiency and accuracy of cell indexing.
For orthorhombic systems, all six axis permutations of $a$, $b$, and $c$ are explored to resolve axis-setting ambiguity. For monoclinic systems, permutations of $a$ and $c$ are also considered.

\vspace{2mm}\noindent
\textit{Stage 2.2 Solving the Cell from an Unknown Space Group.} When the space group is unknown, we developed a \textit{SmartCellSolver} routine that sweeps the base space groups representing the 15 Bravais lattices from highest to lowest symmetry (from cubic face-centered to monoclinic primitive) and then checks the feasibility of each space group within the same Bravais lattice family. To accelerate the search, we also allow early termination to skip low-symmetry cases if good solutions are found in high-symmetry Bravais lattices. Similar to the single-space-group mode, the final solutions are ranked jointly by $\chi^2_\text{cell}$, $N_m$, and unit-cell volume ($V$).

\vspace{2mm}\noindent
\textit{Stage 2.3 Cell Consolidation and WP Assignment.} To reduce the search space, we merge symmetry-equivalent cells using a 5\% tolerance on each lattice parameter and retain at most the top 10 candidate solutions per space group. For each consolidated (SPG, Cell) pair, we enumerate all valid Wyckoff-position (WP) assignments whose site multiplicities reproduce the target stoichiometry within the predicted density bounds. In general, each (SPG, Cell) pair can yield multiple feasible WP assignments.

To further constrain the search, we apply cutoff criteria to limit enumeration depth. These empirically chosen constraints (maximum volume, maximum number of atoms per primitive unit cell) are effective under the assumption that lower-complexity solutions are generally more physically plausible. For each WP combination, we precompute the number of degrees of freedom (DOFs). For the $i$-th WP combination, the $j$-th DOF corresponds to a free fractional coordinate, and its grid complexity (DOF$_{ij}$) is estimated from the cell dimensions ($a$, $b$, $c$) relative to a target spatial resolution (e.g., 1.0~\AA). Because exhaustive sampling over all DOFs can be prohibitively expensive, we estimate the number of trial grids ($N_t$) empirically as
\begin{equation}
N_t = \sum^\text{WP comb}_i \min \left(\prod_j \text{DOF}_{ij},~ \sum_j \text{DOF}_{ij}\right).
\end{equation}

Finally, each (SPG, Cell) candidate is reranked to favor smaller $\chi^2_\text{cell}$, lower $N_m$, smaller $N_t$, and smaller unit-cell volume ($V$), using a composite priority score ($S$):
\begin{equation}
S_i = (\chi^{2, i}_\text{cell})^{0.4}
\left(
\frac{N_t^i}{\min(N_\text{t})}
\cdot
\frac{V^i}{\min(V)}
\cdot
\frac{N_m^i + 1}{\max(N_m) + 1}
\right)^{0.5},
\end{equation}

where the exponents are empirically selected and can be adjusted in future refinements. This procedure produces a ranked list of candidate solutions, as illustrated in the middle panel of Fig.~\ref{fig:workflow}.

\vspace{2mm}\noindent
\textbf{Stage 3: Atomic Structure Determination.}
Given the ranked list of (SPG, Cell, WP) candidates, we proceed to determine the atomic structure by systematically sampling and evaluating candidate configurations.

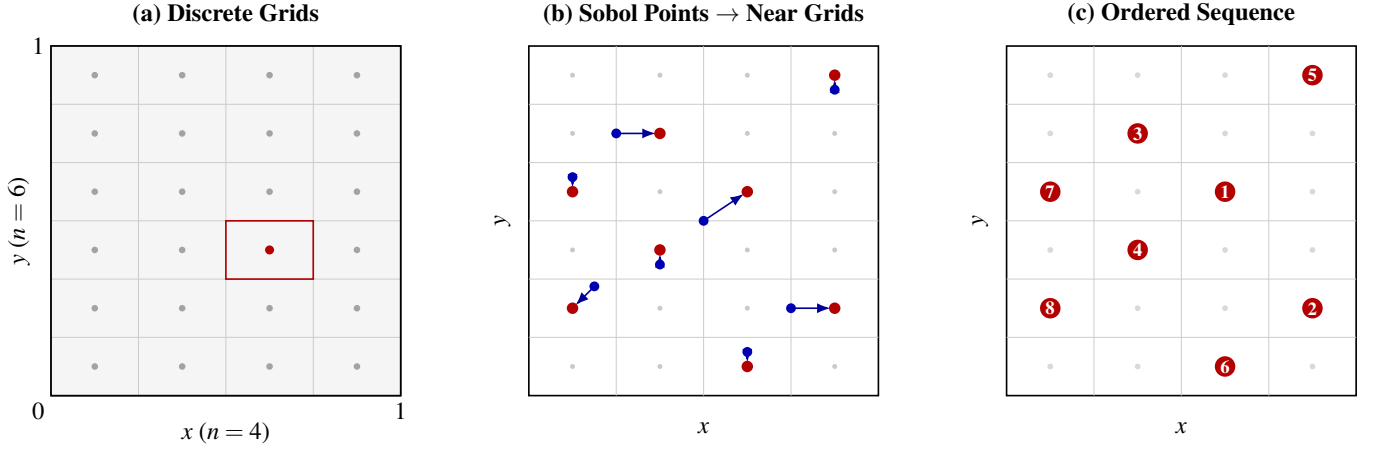
\begin{figure*}[htbp]
\centering
\resizebox{\linewidth}{!}{%
\begin{tikzpicture}[
  >=Latex,
  font=\large,
  box/.style={draw=black!70, thick, rounded corners=2pt},
]

  \colorlet{binfill}{gray!8}
  \colorlet{gridline}{gray!40}
  \colorlet{centerc}{black!60}
  \colorlet{snapc}{red!70!black}
  \colorlet{sobolc}{blue!70!black}
  \colorlet{arrowc}{blue!60!black}

  \begin{scope}[shift={(0,0)}]
    \node[anchor=south] at (3,6.25) {\textbf{(a) Discrete Grids}};
    \draw[thick] (0,0) rectangle (6,6);

    \foreach \i in {0,...,3} {
      \foreach \j in {0,...,5} {
        \fill[binfill] ({6*\i/4},{6*\j/6}) rectangle ({6*(\i+1)/4},{6*(\j+1)/6});
      }
    }
    \foreach \i in {1,...,3}
      \draw[gridline, thin] ({6*\i/4},0) -- ({6*\i/4},6);
    \foreach \j in {1,...,5}
      \draw[gridline, thin] (0,{6*\j/6}) -- (6,{6*\j/6});
    \draw[thick] (0,0) rectangle (6,6);

    \foreach \i in {0,...,3} {
      \foreach \j in {0,...,5} {
        \fill[centerc!60] ({6*(\i+0.5)/4},{6*(\j+0.5)/6}) circle (1.6pt);
      }
    }

    \node[below] at (3,-0.35) {$x$ ($n=4$)};
    \node[rotate=90] at (-0.55,3) {$y$ ($n=6$)};
    \node[below left] at (0,0) {$0$};
    \node[below] at (6,0) {$1$};
    \node[left] at (0,6) {$1$};

    \draw[snapc, thick] (3,2) rectangle (4.5,3);
    \fill[snapc] (3.75,2.5) circle (2.2pt);
  \end{scope}

  \begin{scope}[shift={(8.2,0)}]
    \node[anchor=south] at (3,6.25) {\textbf{(b) Sobol Points $\to$ Near Grids}};
    \draw[thick] (0,0) rectangle (6,6);

    \foreach \i in {1,...,3}
      \draw[gridline, thin] ({6*\i/4},0) -- ({6*\i/4},6);
    \foreach \j in {1,...,5}
      \draw[gridline, thin] (0,{6*\j/6}) -- (6,{6*\j/6});

    \foreach \i in {0,...,3} {
      \foreach \j in {0,...,5} {
        \fill[centerc!35] ({6*(\i+0.5)/4},{6*(\j+0.5)/6}) circle (1.2pt);
      }
    }

    \newcommand{\snaparrow}[4]{%
      \draw[arrowc, thick, ->, shorten >=2pt, shorten <=2pt]
        ({6*#1},{6*#2}) -- ({6*#3},{6*#4});
      \fill[sobolc] ({6*#1},{6*#2}) circle (2.4pt);
      \fill[snapc]  ({6*#3},{6*#4}) circle (2.8pt);
    }

    \snaparrow{0.50}{0.50}{0.625}{0.5833}
    \snaparrow{0.75}{0.25}{0.875}{0.25}
    \snaparrow{0.25}{0.75}{0.375}{0.75}
    \snaparrow{0.375}{0.375}{0.375}{0.4167}
    \snaparrow{0.875}{0.875}{0.875}{0.9167}
    \snaparrow{0.625}{0.125}{0.625}{0.0833}
    \snaparrow{0.125}{0.625}{0.125}{0.5833}
    \snaparrow{0.1875}{0.3125}{0.125}{0.25}

    \node[below] at (3,-0.35) {$x$};
    \node[rotate=90] at (-0.45,3) {$y$};
  \end{scope}

  \begin{scope}[shift={(16.4,0)}]
    \node[anchor=south] at (3,6.25) {\textbf{(c) Ordered Sequence}};
    \draw[thick] (0,0) rectangle (6,6);

    \foreach \i in {1,...,3}
      \draw[gridline, thin] ({6*\i/4},0) -- ({6*\i/4},6);
    \foreach \j in {1,...,5}
      \draw[gridline, thin] (0,{6*\j/6}) -- (6,{6*\j/6});

    \foreach \i in {0,...,3} {
      \foreach \j in {0,...,5} {
        \fill[centerc!25] ({6*(\i+0.5)/4},{6*(\j+0.5)/6}) circle (1.4pt);
      }
    }

    \foreach \x/\y/\lab in {
        0.625/0.5833/1,
        0.875/0.25/2,
        0.375/0.75/3,
        0.375/0.4167/4,
        0.875/0.9167/5,
        0.625/0.0833/6,
        0.125/0.5833/7,
        0.125/0.25/8%
      } {
        \fill[snapc] ({6*\x},{6*\y}) circle (5pt);
        \node[font=\bfseries, white] at ({6*\x},{6*\y}) {\lab};
      }
    \node[below] at (3,-0.35) {$x$};
    \node[rotate=90] at (-0.45,3) {$y$};
  \end{scope}

\end{tikzpicture}%
}
\caption{A cartoon illustration of how Quasi-random sampling (QRS) re-orders the uniform grids. (a) a uniform grid with a resolution of ($n_x$=4 and ($n_y$=6) in a 2D space; (b) the overlay of first eight low-discrepancy Sobol points (blue) and the uniform grids (red); (c) the first eight selected uniform grids according to the Sobol sequence. In this QRS approach, it does not visit all (4$\times$6=24) cells, but progressively select the most important grid points until it hits the desired results.}
\label{fig:QRS}
\end{figure*}

\vspace{2mm}\noindent
\textit{Stage 3.1 Quasi-Random Sampling (QRS).} For each ranked (cell, SPG, WP) triplet, candidate atomic configurations are generated by assigning (0, 1) fractional coordinates to the free Wyckoff parameters within the unit cell. In contrast to conventional stochastic optimization approaches~\cite{Oganov-NRM-2019}, we adopt a deterministic QRS strategy to ensure full reproducibility of the search process~\cite{joe2008constructing}. Specifically, we construct anisotropic discrete grids from the previously introduced DOF$_{ij}$ values along each crystallographic direction. Rather than traversing these grids in lexicographic order, the grid points are reordered according to a low-discrepancy sequence (e.g., Sobol or Halton), which maximizes global coverage of configurational space from the earliest stages of the search. This strategy is graphically illustrated in Fig. \ref{fig:QRS}. As a result, even a modest number of sampled configurations can efficiently explore diverse regions of the high-dimensional coordinate space. Moreover, the progressive nature of low-discrepancy sequences allows the search to be refined systematically as additional samples are generated, enabling the configurational space to be explored in an increasingly uniform and hierarchical manner. This property is particularly advantageous when structural relaxations are computationally expensive and only a limited number of candidate structures can be evaluated.

\vspace{2mm}\noindent
\textit{Stage 3.2 Geometry Optimization and Pre-screening.} Each trial structure is relaxed using the MACE universal neural-network force field~\cite{Batatia2022mace} without changing the cell parameters via ASE~\cite{ase}. After relaxation, structures with residual forces exceeding 0.5~eV/\AA~ or stresses exceeding 0.3~GPa are considered infeasible and discarded. Only well-relaxed structures are retained and evaluated by comparing their PXRD patterns with the observed experimental data using the normalized cross-correlation function~\cite{de2001generalized, habermehl2014structure}. This combined energy and PXRD similarity filtering effectively reduces computational cost for subsequent refinement steps.

\begin{table*}[htbp]
\centering
\caption{Summary of candidate structure solutions for $I4/mmm$ Eu$_3$Al$_2$O$_7$ ranked by the priority score.}
\label{tab:candidates}
\setlength{\tabcolsep}{8pt}
\begin{tabular}{cclcccccc}
\hline
Pair & Space Group & Cell (\AA) &\#WP Comb.& Volume (\AA$^3$) & $\chi^2_\text{cell}$ & $N_m$ & $N_t$  & Priority Score \\
\hline
1  &   47  &  [ 3.756,  19.820,   3.755] &1&      279.6&      0.0003&     4&      30&     0.020\\
2  &   47  &  [19.820,   3.756,   3.755] &1&      279.6&      0.0003&     4&      30&     0.020\\
3  &   66  &  [ 5.309,   5.311,  19.820] &1&      558.9&      0.0002&     2&      40&     0.022\\
4  &   68  &  [ 5.309,   5.311,  19.820] &1&      558.9&      0.0002&     2&      44&     0.023\\
5  &   68  &  [19.820,   5.309,   5.311] &1&      558.9&      0.0002&     2&      44&     0.023\\
6  &   42  &  [19.820,   5.309,   5.311] &1&      558.9&      0.0002&     2&      69&     0.029\\
7  &   59  &  [19.820,   3.756,   3.755] &1&      279.6&      0.0003&     4&      65&     0.030\\
8  &   66  &  [19.820,   5.309,   5.311] &1&      558.9&      0.0002&     2&      82&     0.032\\
\textbf{9}  &   \textbf{139} &  \textbf{[ 3.755,  19.811, 19.811]} &\textbf{1}&      \textbf{279.4}&      \textbf{0.0004}&     \textbf{3}&      \textbf{76}&     \textbf{0.032}\\
10 &   68  &  [ 5.309,  19.820,   5.311] &1&      558.9&      0.0002&     2&      86&     0.032\\
$\cdots$&$\cdots$&$\cdots$&$\cdots$&$\cdots$&$\cdots$&$\cdots$&$\cdots$ & $\cdots$ \\
158&   57&    [17.285,   7.022,   4.955] &2&      601.4&      0.0335&    27&     385&     1.625\\
\hline
\end{tabular}
\end{table*}

\begin{figure*}[htbp]
\vspace{3mm}
    \centering
\includegraphics[width=0.99\textwidth]{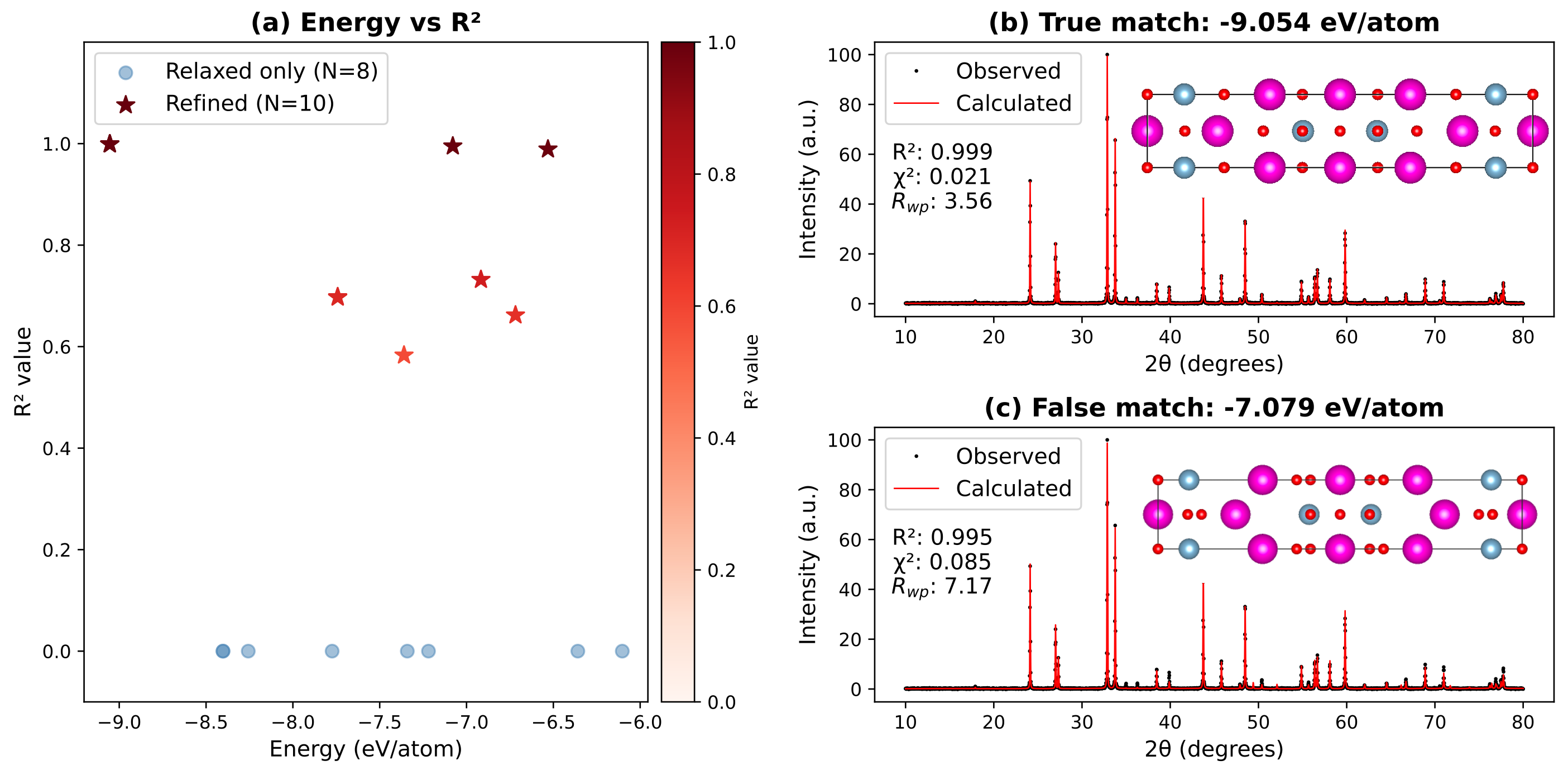}
    \caption{Application of Ab-PXRD-Solver to the solution of the $I4/mmm$ Eu$_3$Al$_2$O$_7$ structure. (a) Energy vs. $R^2$ relationship for all sampled structures; (b) and (c) crystal structures corresponding to true and false matches as indicated by the energy ranking.}
    \label{fig3}
    \vspace{-2mm}
\end{figure*}  

\vspace{2mm}\noindent
\textit{Stage 3.3 Refinement, Termination \& Restart.}
Relaxed structures exhibiting high PXRD similarity are selected for Rietveld refinement using GSAS-II~\cite{toby2013gsas2}. A solution is accepted and the search terminates when $R^2 \geq 0.95$ and $\chi^2 \leq 0.12$. The final output includes structural details, MACE energies, and refinement metrics such as $R_\mathrm{wp}$ for further evaluation (see the lower panel plot of Fig.~\ref{fig:workflow}). If no solution meets the acceptance criteria, the search continues with the next ranked (SPG, Cell, WP) candidate until all candidates are exhausted or a solution is found. If no acceptable solution is reported, an expert user can analyze the historical results and decide whether to relax the acceptance criteria, manually adjust the peak assignment, or expand the search space by considering additional (SPG, Cell) candidates with a larger volume cutoff or more trial grids for a given (SPG, Cell, WP) pair.

\vspace{3mm}\noindent
\textbf{\large{3. Example Usage}}\\
Fig.~\ref{fig3} presents the results of solving the $I4/mmm$ Eu$_3$Al$_2$O$_7$ crystal structure with and without prior space-group information. When the correct space group (No.~139) is provided, the workflow identifies only one valid (SPG, Cell) candidate pair, with one possible WP combination and a total of 76 trial grids. In contrast, when the space group is treated as unknown, the workflow estimates 158 (SPG, Cell) candidate pairs, corresponding to 33,194 trial structures (see Table~\ref{tab:candidates}). Despite this substantially expanded search space, the prioritization strategy still ranks the ground-truth solution favorably (9/158), enabling early termination once a promising solution is found; consequently, the computational effort increases only modestly. In this case, none of the candidates ranked 1--8 yields a valid structure. For space group 139, the 76 trial grids generate 18 successfully relaxed structures, of which 10 are selected for Rietveld refinement, yielding three ground-truth hits with $R^2$=0.999, $R_\text{wp}$=3.56, and $\chi^2$=0.021 (Fig.~\ref{fig3}b). The correctness of the obtained structure is further supported by its identification as the lowest-energy configuration at -9.054~eV/atom.


\begin{figure*}[htbp]
    \centering
\includegraphics[width=0.99\textwidth]{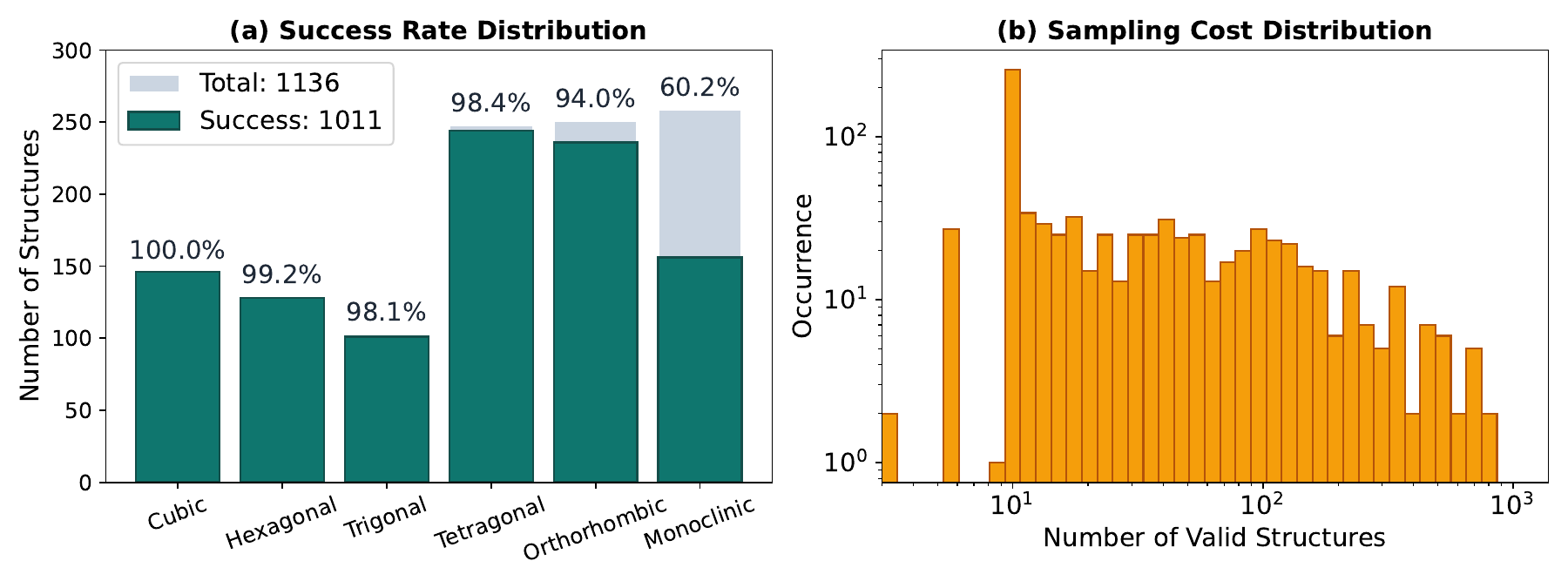}
\vspace{-3mm}
    \caption{Application of Ab-PXRD-Solver to 1136 hard examples from previous studies. (a) Distribution of success rates; (b) Distribution of sampled valid structures in each successful examples.}
    \label{fig4}
\end{figure*}  

It is important to note that many previous approaches rely primarily on either $R^2$ or $R_\mathrm{wp}$ as the sole metric for evaluating PXRD refinement quality~\cite{guo2025ab, li2025powder, lu2026unified}. However, a structure with an excellent diffraction fit may still correspond to a metastable or physically unreasonable configuration with relatively high energy. As shown in Fig.~\ref{fig3}c, we also identified a structure with good refinement quality ($R^2$=0.995 and $R_\mathrm{wp}$=7.17); however, its energy is much higher at -7.079~eV/atom.
Therefore, considering both diffraction agreement and energetic stability provides a more robust criterion for identifying physically meaningful crystal structures.

Unlike recent AI-generative approaches that aim to solve crystal structures within seconds~\cite{choudhary2025diffractgpt, guo2025ab, li2025powder, lu2026unified}, the present solver typically requires longer computation times, ranging from several minutes to several hours on a single CPU, but offers substantially greater physical interpretability and reliability. First, it samples a larger number of candidate structures to better ensure that the final solution corresponds to a physically plausible, low-energy configuration. Second, it directly generates symmetrized crystal structures that strictly satisfy the target space-group symmetry, thereby avoiding the need for post hoc symmetry corrections.

\vspace{3mm}\noindent
\textbf{\large{4. Test on a large Dataset}}\\
To demonstrate the effectiveness of our approach, we focus on the previously failed examples reported in Table~\ref{tab:stat2}. As summarized in Fig.~\ref{fig4}a, after excluding several 2D monolayered and triclinic cases, this dataset contains 146 cubic, 129 hexagonal, 103 trigonal, 248 tetragonal, 251 orthorhombic, and 259 monoclinic structures. For these 1136 structures, we performed fully automated tests using \textit{SmartCellSolver} to infer the space group. The resulting success rates are 100\% (cubic), 99.2\% (hexagonal), 98.1\% (trigonal), 98.4\% (tetragonal), 94.0\% (orthorhombic), and 60.1\% (monoclinic). In most high-symmetry cases, the correct solution is ranked within the top 10 candidates, allowing the search to terminate after evaluating only a few candidates. For low-symmetry cases, however, the search space becomes substantially larger, consistent with the expected increase in search complexity. The failure cases are often associated with two reasons: (i) the correct (Cell, SPG) solution cannot be found, and (ii) the ground truth solution is ranked too low and it exceeds the maximum number of candidates that can be evaluated within the computational budget. 



Fig.~\ref{fig4}b shows the distribution of number of successfully relaxed structures for all 1136 cases. For a lot of cases (447 out of 1136), the workflow successfully relaxes no more than 20 candidate structures due to the early termination mechanism. 
For a few challenging cases, only a limited number of structures can be successfully relaxed, which may require more extensive sampling or refinement to identify the correct solution. Currently, we simply follow the pre-assigned order and resolution to sample the QRS grid of each (SPG, Cell, WP) solution. For very complicated cases, it may be advantageous to develop a separate reinforcement-learning-based ranking strategy to adaptively prioritize the most promising candidates for further sampling and refinement.

Finally, it is important to emphasize that the success rates reported here are based on fully automated runs without manual intervention. Unlike purely AI-based solvers, our workflow is not a black-box optimization procedure. When a structure search fails to produce a satisfactory fit, experienced materials researchers can inspect intermediate results to diagnose the underlying causes, adjust key parameters (e.g., peak-selection thresholds, candidate-cell tolerances, search-space constraints, or denser QRS grid points), and then perform a more extensive search over cell parameters and atomic configurations. This human-interpretable design enables iterative refinement and offers greater transparency than fully automated generative approaches.

\vspace{3mm}\noindent
\textbf{\large{5. Discussion and Outlook}}\\
Having demonstrated its efficiency in solving commonly failed examples by two previously developed AI-generative models, we emphasize that the key strength of the present approach is that it is nearly \textit{ab initio} in spirit while remaining computationally tractable for complex inorganic materials. Compared with many previous crystal-structure prediction and PXRD-solution methods, the present framework is capable of treating structures well beyond the conventional $\sim$20-atom limitation commonly encountered in exhaustive searches. At the same time, the workflow maintains high fidelity in the final solution by simultaneously considering diffraction agreement, crystallographic symmetry, and energetic plausibility.

Methodologically, the core problem is formulated as two sequential optimization tasks: (i) identifying candidate lattice and symmetry by enumerating the mappings between the observed PXRD peaks and finite (hkl) indices in discrete space, and (ii) optimizing atomic arrangements over combinations of Wyckoff sites in continuous space using a deterministic quasi-random sampling technique. This decomposition substantially reduces the complexity of the search space while preserving crystallographic rigor. In addition, physically motivated cutoff criteria, such as upper bounds on hkl indices and unit-cell volume, are applied throughout the workflow to ensure practical convergence and enable early termination once sufficiently promising solutions have been identified.

AI-based models are incorporated selectively into tasks for which pattern recognition is especially beneficial, such as peak identification and candidate filtering, whereas the core crystallographic search remains grounded in physics-based solvers and diffraction theory. Equally important, the recent development of the universal interatomic potential MACE plays a key role in enabling rapid yet physically meaningful energy evaluation across chemically diverse systems. The use of other potentials, such as \texttt{MatterSim}~\cite{yang2024mattersim}, \texttt{MACE-OFF}~\cite{kovacs2025mace}, and \texttt{UMA}~\cite{wood2025family}, may further improve accuracy and facilitate efficient structure determination for organic crystals and metal-organic framework (MOF) systems.

Despite these advantages, substantial room for improvement remains. For highly complex, low-symmetry systems with large unit cells and many degrees of freedom, the number of candidate solutions can become extremely large. In such cases, the correct solution may not rank near the top during the early stages of the search, potentially requiring more extensive sampling and refinement. Future improvements may include more adaptive ranking strategies through reinforcement learning, tighter integration between diffraction fitting and energetic evaluation, and the incorporation of uncertainty-aware AI models to further enhance search efficiency and robustness for challenging materials systems.

\vspace{3mm}\noindent
\textbf{\large{Data availability}}\\
The complete list of 1136 structures and the solutions are available in \url{https://mmi.charlotte.edu/ab_pxrd_solver}.

\vspace{3mm}\noindent
\textbf{\large{Code availability}}\\
The \texttt{Ab-PXRD-Solver} source code, instructions, as well as scripts used to calculate the results of this study, are available in \url{https://github.com/MaterSim/Ab-PXRD-Solver}.

\vspace{3mm}\noindent
\textbf{\large{Acknowledgments}}\\
This research was sponsored by the U.S. Department of Energy, Office of Science, Office of Basic Energy Sciences, and the Established Program to Stimulate Competitive Research (EPSCoR) under the DOE Early Career Award No. DE-SC0024866, the UNC Charlotte's seed grant for data science. The computing resources are provided by ACCESS (TG-MAT230046). 

\vspace{3mm}\noindent
\textbf{\large{Author contributions}}\\
Q.Z. conceived the idea and initiated the framework. Q.Z. and H.X. supervised this project. All authors implemented the code, analyzed the results and contributed to manuscript writing.

\vspace{3mm}\noindent
\textbf{\large{Competing interests}}\\
All authors declare that they have no conflict of interest.

\vspace{3mm}\noindent
\textbf{\large{Additional information}}\\
Supplementary information The online version contains
supplementary material available at
https://doi.org/****/*****

\vspace{3mm}\noindent
\textbf{\large{References}}

\bibliography{ref}

\end{document}